\documentclass{appolb}
\usepackage{epsfig}

\begin{document}
\title{Charge conservation effects for
high order fluctuations
\thanks{Presented at Critical Point and Onset of Deconfinement 2016 (CPOD2016).} 
}
\author{Viktor Begun
 \address{Faculty of Physics, Warsaw University of Technology, Koszykowa 75, 00-662 Warsaw, Poland }
}
\maketitle
\begin{abstract}
The exact charge conservation significantly impacts multiplicity
fluctuations. The result depends strongly on the part of the
system charge carried by the particles of interest.
Along with the expected suppression of fluctuations for large
systems, charge conservation may lead to negative skewness or
kurtosis for small systems.
\end{abstract}

 \vspace{0.3cm}

The STAR collaboration observes the non-monotonous behavior of the
net-proton normalized kurtosis~\cite{Thader:2016gpa}. This might
be an indication of the critical point of strongly interacting
matter. The NA61/SHINE collaboration performs the system size and
energy scan in order to find the critical point and study it's
properties~\cite{MajaMackowiak-PawlowskafortheNA61/SHINE:2016aci}.
Electric charge Q, baryon number B, and strangeness S are
conserved exactly in strong interactions. This may change the
expected background fluctuations in high energy
collisions\footnote{The grand canonical statistical ensemble
(GCE), where charges are conserved on average, is equivalent to
the canonical ensemble (CE), where charges are conserved exactly,
only for mean multiplicities and only for large systems.
Multiplicity fluctuations in CE and GCE for both small and large
systems are different~\cite{Begun:2004zb}.}.
%

For illustration purposes the system with only one conserved
charge in CE Hadron Gas model for primary particles (no resonance
decays) is considered in the system with the total charge Q=2.
The straightforward calculations using~\cite{Begun:2004zb} allow
to obtain fluctuations for different amount of charge carried by
the particles, $k=z_j/z$, where $z=\sum_jz_j$, and $j$ is the type
of hadron, and $z_j$ is the one particle partition
function~\cite{Begun:2004zb}.
The k=0.95 means that the particles contain 95\% of the electric
charge Q=2 of the system, which may correspond to pions created in
p+p reactions. Let's also consider the case k=0.5, in order to
check the importance of this parameter.
The normalized skewness, $S\cdot\sigma=m_3/m_2$, where $m_i$ are
central moments, and normalized kurtosis,
$\kappa\cdot\sigma^2=(m_4-3m_2^2)/m_2$, are considered. They are
calculated for plus, minus, net, and total number of charged particles. 
The results are shown as the function of the mean number of
neutral particles $\langle N_0\rangle$, which is the same in GCE
and CE.
\begin{figure}
\includegraphics[width=\textwidth]{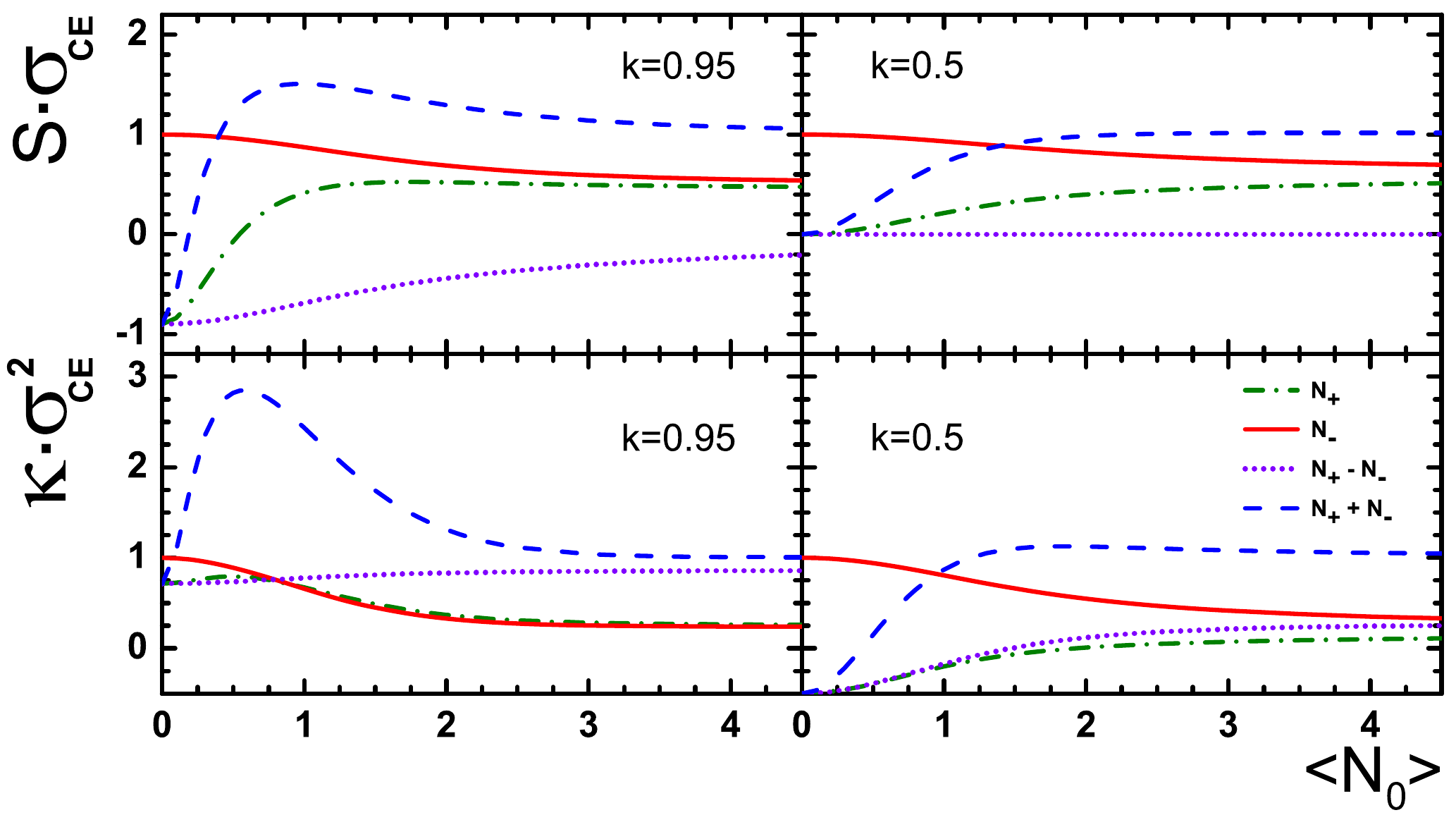}
\caption{The normalized skewness (top row) and the normalized
kurtosis (down row) for different amount $k$ of the system charge
carried by the particles of interest.}
\end{figure}
%
%
One can see that all observables have a non-trivial behavior in
CE\footnote{GCE values of $S\cdot\sigma$ and $\kappa\cdot\sigma^2$
are either 1 for Poisson, or 0 for Gauss distribution.}.
Since $Q=N_+-N_-=2>0$, fluctuations of $N_+$ and $N_++N_-$ are
suppressed for small systems, while $N_-$ fluctuate as by Poisson.
For large systems, $\langle N_0\rangle\gg Q$, "+" and "-" become
equivalent. The net-charge $N_+-N_-$ is always special. The
$S\cdot\sigma_{_{\rm CE}}$ can be negative for $k=0.95$, and
$\kappa\cdot\sigma^2_{_{\rm CE}}$ can be negative for $k=0.5$,
while they have positive values in other cases. Thus, the
calculations and interpretation of high moments should be
performed with a great care.


{\bf Acknowledgments:} The author thanks to M.I. Gorenstein and
M.~Mac\-ko\-wiak-Pawlowska for discussions. This work was
supported by Polish National Science Center grant No.
DEC-2012/06/A/ST2/00390.

\end{document}